\documentclass[sigconf,authorversion, nonacm]{acmart}

\AtBeginDocument{%
  }

\usepackage{subcaption}
\usepackage{multirow}
\usepackage{siunitx}
\usepackage{enumitem}

\title[NeuroTouch: A Neuromorphic Optical Tactile Sensor for Real-Time Gesture Recognition]{From Soft Materials to Controllers with NeuroTouch: A Neuromorphic Tactile Sensor for Real-Time Gesture Recognition}

\author{Victor Hoffmann}
\email{Victor.Hoffmann@sony.com}
\affiliation{%
  \institution{Sony Semiconductors Solutions, \\ Sony Europe B.V.,\\ Europe Imaging, Sensing and Perception Center}
  \city{Zurich}
  \country{Switzerland}}

\author{Federico Paredes-Valles}
\email{Federico.Paredes-Valles@sony.com}
\affiliation{%
  \institution{Sony Semiconductors Solutions, \\ Sony Europe B.V.,\\ Europe Imaging, Sensing and Perception Center}
  \city{Zurich}
  \country{Switzerland}}

\author{Valentina Cavinato}
\email{Valentina.Cavinato@sony.com}
\affiliation{%
  \institution{Sony Semiconductors Solutions, \\ Sony Europe B.V.,\\ Europe Imaging, Sensing and Perception Center}
  \city{Zurich}
  \country{Switzerland}}

\renewcommand{\shortauthors}{Hoffmann et al.}
\begin{document}

\begin{abstract}
This work presents NeuroTouch, an optical-based tactile sensor that combines a highly deformable dome-shaped soft material with an integrated neuromorphic camera, leveraging frame-based and dynamic vision for gesture detection. Our approach transforms an elastic body into a rich and nuanced interactive controller by tracking markers printed on its surface with event-based methods and harnessing their trajectories through RANSAC-based techniques. To benchmark our framework, we have created a \SI{25}{\minute} gesture dataset, which we make publicly available to foster research in this area. Achieving over \SI{91}{\percent} accuracy in gesture classification, a \SI{3.41}{\milli\meter} finger localization distance error, and a \SI{0.96}{\milli\meter} gesture intensity error, our real-time, lightweight, and low-latency pipeline holds promise for applications in video games, augmented/virtual reality, and accessible devices. This research lays the groundwork for advancements in gesture detection for vision-based soft-material input technologies. Dataset: Coming Soon, Video: Coming Soon
\end{abstract}


\begin{teaserfigure}
\centering
\includegraphics[width=\textwidth]{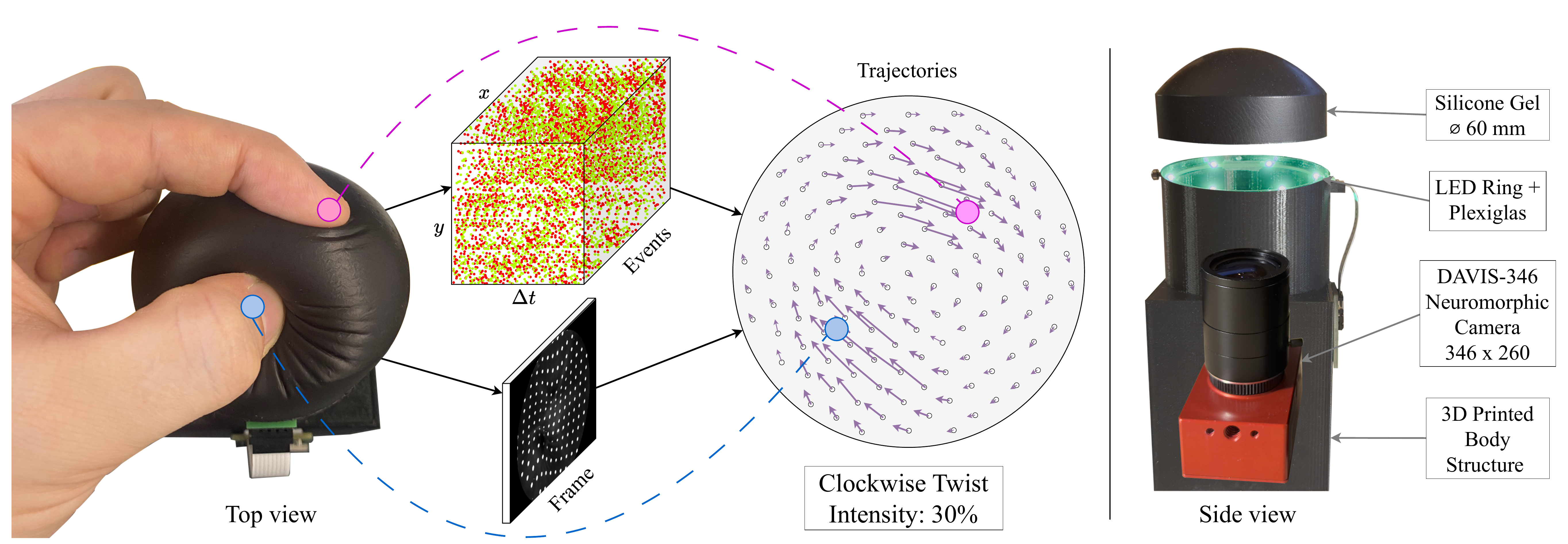}
\caption{(Left) NeuroTouch is a vision-based soft-material controller that utilizes both frame-based and event-based visual processing. By analyzing the trajectories of markers printed on the surface of the soft material, the pipeline estimates in real time the user's gesture type, finger localization, and gesture intensity. (Right) System overview. Silicone gel is shown detached for clarity. The DAVIS-346 camera is placed inside the body structure.}
\Description{Left image is fully described in caption. On the right image, a DAVIS-346 camera is placed in a 3D body structure below the silicone gel to see the markers. An LED ring is placed on top of the structure between the camera and the gel in order to have enough brightness.}
\label{fig:teaser}
\end{teaserfigure}


\maketitle

\section{Introduction}
%Motivation: Why is this problem important?
%Problem Statement: What specific problem is being addressed?
%Contributions: Clear enumeration of the main contributions of the paper.
%Structure of the Paper: An outline of the sections to follow.

Interactive devices form the bridge between users and digital environments, with applications spanning from video games, augmented reality (AR), virtual reality (VR), and beyond. However, despite their pivotal role, current interactive devices often impose limitations on users, particularly gamepads and rigid controllers. These devices, while robust and precise, can lack the expressiveness and ergonomic adaptability necessary for natural and intuitive interactions. Moreover, their rigidity and conventional designs frequently fail to accommodate users with impaired hand functions, limiting accessibility \cite{game_accessibility}.

This paper introduces NeuroTouch, a vision-based soft-material controller designed to address these challenges. Built with a highly deformable silicon gel, this interactive device enables intuitive and ergonomic tactile interactions through multi-finger gesture detection. By leveraging a neuromorphic camera for real-time tracking of markers on the gel, the system maintains high performance even in high-speed scenarios. Our gesture detection pipeline offers precise finger position tracking, accurate gesture type classification, and robust intensity estimation. Our primary contributions are as follows:
\begin{itemize}[label=\textbullet, left=10pt]
    \item We present NeuroTouch, a soft-material controller integrating optical-based tactile sensing via a neuromorphic camera (combining event-based and frame-based imaging).
    \item We propose a CPU-only, lightweight and high-frequency gesture detection pipeline that enables precise localization of finger positions, classification of gesture types, and estimation of gesture intensity using only the camera's input. 
    \item We introduce a gesture detection dataset designed to evaluate the performance of our pipeline, providing benchmarks for prediction accuracy and runtime efficiency. This dataset is freely available to promote research in vision-based soft-material controllers.
\end{itemize}

\section{Related Work}
\label{related_work}

Over the years, interactive soft-material controllers have significantly evolved, enabling novel haptic and user interaction applications. Early systems like SOFTii \cite{SOFTii} integrate shape-changing features and interactive feedback to simulate textures, while FoamSense \cite{FoamSense} and Skin-On Interfaces \cite{Skin-On-Interfaces} advance pressure-sensitive materials and tactile interactions. However, many sensors remain limited to 2D or minimal deformation responses. Approaches like deForm \cite{deForm} addresses this issue with structured light for 2.5D feedback but faces constraints in scalability and deformation range.

Optical-based tactile sensors provide a promising alternative. Originally developed with systems such as \cite{tactile_sensor_original}, these sensors combine deformable surfaces with internal optics, using markers or light patterns \cite{tactile_light_patterns} to detect touch and measure forces. Typically illuminated by LEDs, cameras capture marker displacements, which are processed for specific tasks \cite{GelForce}. These sensors excel in robotics applications, including slip detection \cite{tactile_slip_detection}, force estimation \cite{tactile_sensor_force_estimation}, texture recognition \cite{NeuroTac}, grasp stability \cite{tactile_grasp_adjustments}, object recognition \cite{tactile_recognition}, and pose estimation \cite{tactile_object_pose_estimation}, but often lack high frequency.

Event-based optical tactile sensors overcome these limitations by replacing standard cameras with event-based vision sensors, which asynchronously capture per-pixel positive and negative logarithmic brightness changes, yielding sparse data, high temporal resolution, and low power usage \cite{DVS}. Advances such as \cite{NeuroTac, Evetac, NeuTouch} showcase their potential as high-frequency, low-power solutions but are tailored exclusively for robotic tactile sensing applications.

Beyond robotics, optical-based tactile sensors hold significant potential for human-computer interaction by detecting finger gestures and applied forces. GelForce \cite{GelForce} has pioneered this domain, but is constrained to force field estimation, limited gel deformations, and a low operational frequency due to the charged coupled device camera frame rate. Compact designs such as OneTip \cite{OneTip} extend capabilities to six degrees of freedom but remain confined to single-finger interactions. To our knowledge, no prior research has tackled multi-finger gesture detection on vision-based tactile sensors, whether using frame- or event-based techniques. This gap underscores a lack of datasets, benchmarks, and methodologies optimized for this promising application domain.

\section{System Overview}

\begin{comment}
\begin{figure}[htbp!]
    \centering
    \includegraphics[trim={10pt 100pt 10pt 80pt}, clip, width=0.8\columnwidth]{figures/system_overview.pdf}
    \caption{NeuroTouch overview. The silicone gel is shown detached for clarity. The DAVIS-346 camera is placed inside the body structure.}
    \Description{This figure shows the different parts of NeuroTouch. A DAVIS-346 camera is placed in a 3d body structure below the silicone gel to see the markers. An LED ring is placed on top of the structure between the camera and the gel in order to have enough brightness.}
    \label{fig:overview-sensor}
\end{figure}
\end{comment}

An overview of our tactile sensor is shown in Figure~\ref{fig:teaser}. It is composed of three components usually found in optical-based tactile sensors: a silicone gel, LEDs, and a camera \cite{GelSight}. However, our sensor introduces two notable features that make it stand out.

\begin{figure}[htbp!]
    \centering
    \begin{subfigure}{0.48\columnwidth}
        \centering
        \includegraphics[trim={0pt, 100pt, 0pt, 0pt}, clip, width=\linewidth]{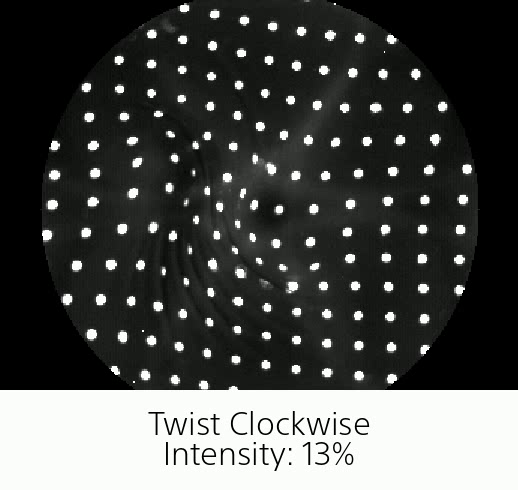}
        \caption{APS frame}
        \label{fig:input_example_frame}
    \end{subfigure}
    \begin{subfigure}{0.48\columnwidth}
        \centering
        \includegraphics[trim={0pt, 100pt, 0pt, 0pt}, clip,width=\linewidth]{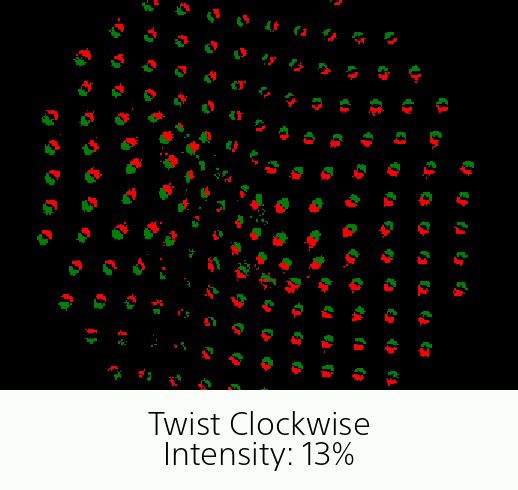}
        \caption{Event frame}
    \end{subfigure}
    \caption{APS frame (a) and EVS events (b) example of a two-finger \textit{Clockwise Twist} gesture. Events are accumulated in a 10 ms frame, with positive brightness changes in green and negative ones in red.} 
    \Description{Fig (a) depicts the APS frame. As fingers press and twist on the gel, they deform its surface and displace the embedded markers. The spatial distribution of events highlights the dynamics of the gesture: since our markers are white and the background is black, positive events slightly precede negative ones during a motion, indicating that the gesture is in the releasing phase.}
    \label{fig:inupt-example}
\end{figure}

The first distinctive feature is the size and shape of the silicone gel. Unlike gels typically used in robotics \cite{GelSight, tactile_robot_finger_1, of_mlp_tactile_sensor}, our gel is relatively large (\SI{60}{\milli\meter} diameter) and has a unique curved shape, resembling a dome. This design accommodates high deformations, multi-finger gestures and facilitates natural interactions. The gel's hardness mimics that of human skin, offering a tactile experience that feels intuitive and organic. The gel surface is made of silicone with a black surface embedded with $177$ white markers, forming a grid-like pattern that aids in precise motion tracking. Each marker is a dot with a diameter of \SI{1}{\milli\meter}, regularly spaced \SI{4}{\milli\meter} apart, making the dilation of the markers small when a finger is applied to the gel. Figure~\ref{fig:inupt-example} illustrates a representation of the markers as viewed from the camera's perspective.

\label{subsec:neuromorphic_camera}
The second feature lies in the use of a neuromorphic camera system that combines a standard Active Pixel Sensor (APS) with an Event-Based Vision Sensor (EVS). High temporal resolution, as provided by the EVS, is essential for tracking marker movements during rapid gestures. Standard cameras, typically operating at \num{25}-\SI{50}{\hertz}, struggle to capture fast marker displacements, leading to motion blur and significant gaps in positional data, which increases the likelihood of tracking errors. While high-speed APS cameras could mitigate these issues, they come at the cost of increased power consumption and can struggle to track many markers in real-time due to computational constraints
\cite{High-Speed-Camera-Tracking}. In contrast, EVS technology processes sparse and near-continuous data streams, enabling high-frequency, reliable, and precise marker tracking even during high-speed movements \cite{evs_vs_frame_tracking_2, evs_vs_frame_tracking_1}. 

Alternatively, relying solely on an EVS presents its challenges. Due to the sensor's differential nature, static scenes generate minimal events, most of which are related to sensor noise \cite{EVS_Survey}. Under these conditions, it is difficult to distinguish genuine markers from noise artifacts, leading to a deterioration in tracking quality over time. Consequently, distinguishing between two static scenarios, such as holding a gesture versus the gel being in a resting position, becomes increasingly difficult.

By combining the strengths of both an EVS and an APS, our approach achieves optimal results, leveraging the high temporal resolution of the EVS for dynamic tracking and the APS for static scene analysis. For NeuroTouch, we have used a DAVIS-346 camera\footnote{More information on the camera can be found here: \url{https://inivation.com/wp-content/uploads/2019/08/DAVIS346.pdf}} which has a resolution of $346\times260$, capturing frames at \SI{25}{Hz} and events at microsecond resolution. We believe that the low resolution of our camera is sufficient for accurately tracking the markers while maintaining low power consumption and achieving fast image processing runtimes.

\section{Problem Statement}
\label{sec:problem_statement}
%Clear definition of the problem, including assumptions and constraints.
%Mathematical formulation (e.g., equations, models, optimization objectives) when applicable.

\begin{figure}[t!]
    \centering
    % Top row
    \begin{subfigure}{0.25\columnwidth}
        \centering
        \includegraphics[width=\linewidth]{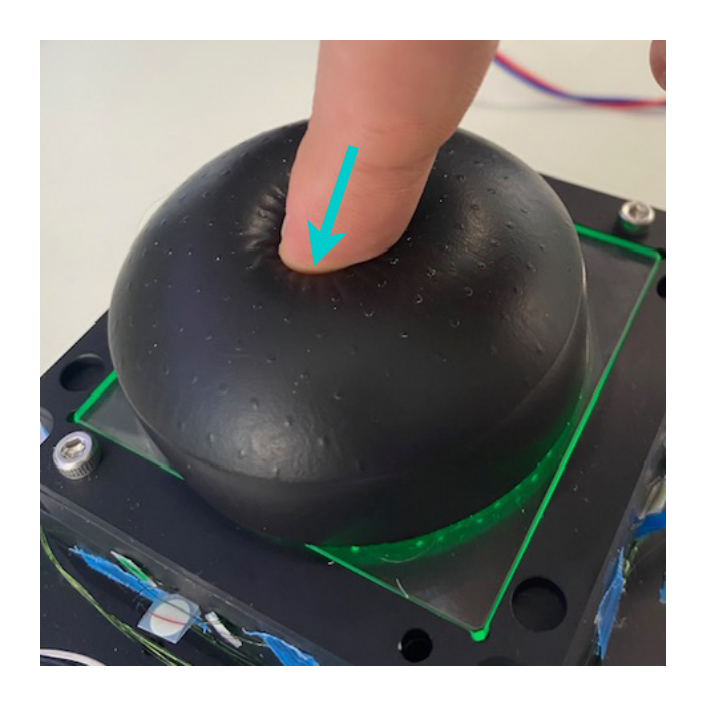}
        \caption{\textit{Push}}
    \end{subfigure}\hspace{-0.5em}
    \begin{subfigure}{0.25\columnwidth}
        \centering
        \includegraphics[width=\linewidth]{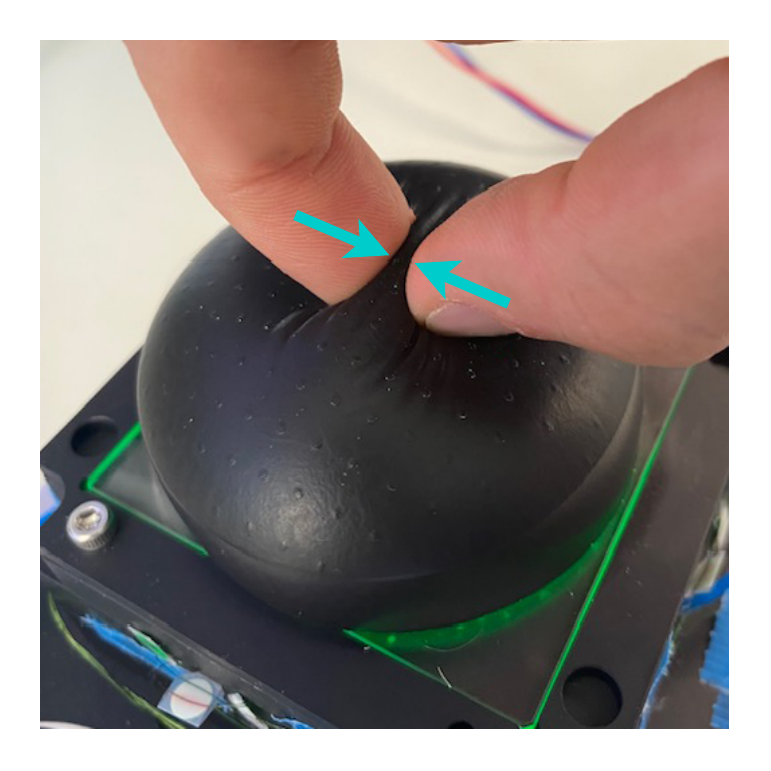}
        \caption{\textit{Pinch}}
    \end{subfigure}\hspace{-0.5em}
    \begin{subfigure}{0.25\columnwidth}
        \centering
        \includegraphics[width=\linewidth]{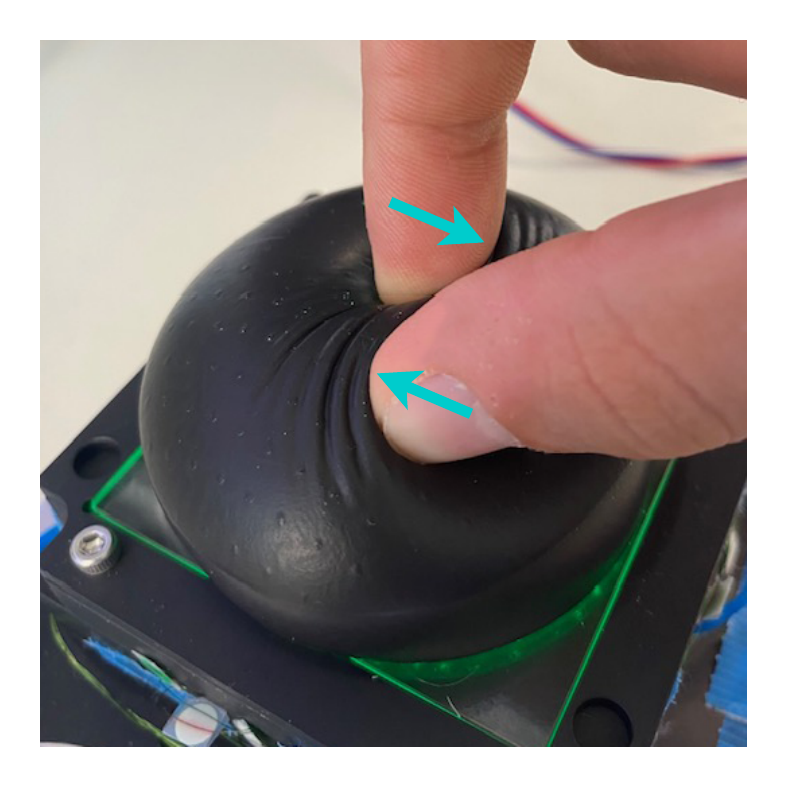}
        \caption{\textit{Twist}\protect\footnotemark}
    \end{subfigure}\hspace{-0.5em}
    \begin{subfigure}{0.25\columnwidth}
        \centering
        \includegraphics[width=\linewidth]{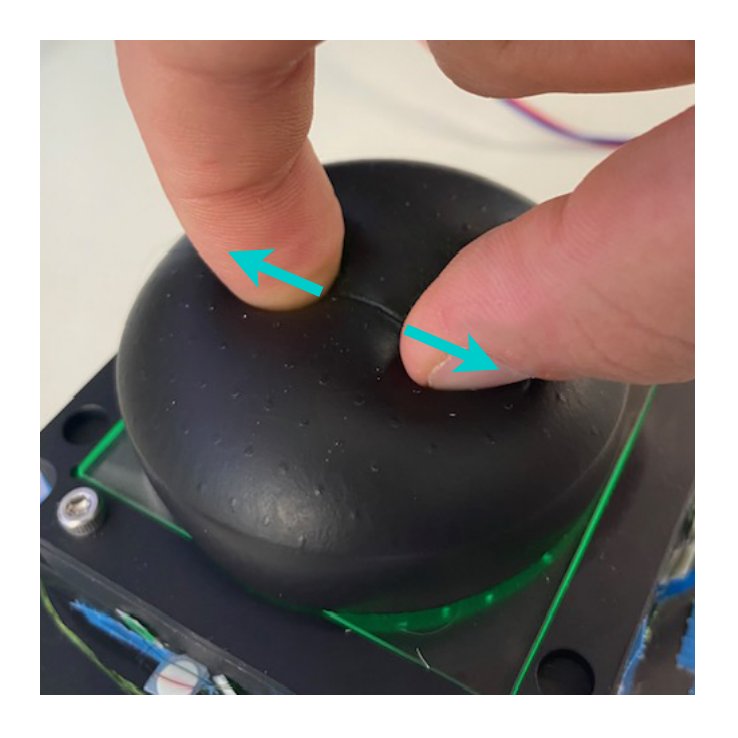}
        \caption{\textit{Zoom}}
    \end{subfigure}
    
    \caption{Illustration of the basic gesture types used to classify the user’s actions on the gel. Arrows indicate the direction of finger movements.}
    
    \Description{Subfigures illustrate four gesture types that a user can apply on the gel: Push, Pinch, Twist and Zoom. The Twist gesture is further split into Clockwise Twist and Counter-Clockwise Twist.}
    \label{fig:gesture_types}
\end{figure}

Gesture detection on optical-based tactile sensors has, to our knowledge, not been explored yet. As such, there are no established benchmarks or universally accepted definitions of tactile gestures on such devices. To address this, we propose a foundational framework for defining and interpreting gestures in this context. Specifically, we define a gesture through three key components: 
\begin{itemize}[label=\textbullet, left=10pt]
    \item \textbf{Localization of contact points}: The positions of the fingers as they touch and interact with the silicone surface. Since a finger covers a finite area rather than a single point, a contact point is defined as the location corresponding to the maximum deformation of the silicone gel caused by a finger's interaction.
    \item \textbf{Gesture type}: The classification of the user's action. In this work, we categorize gestures into five basic types: \textit{Push}, \textit{Pinch}, \textit{Zoom}, \textit{Clockwise Twist}, and \textit{Counter-Clockwise Twist} (cf. Figure~\ref{fig:gesture_types}).
    \item \textbf{Gesture intensity}: A measure of the deformation magnitude of the gel around the fingers performing the gesture, providing a quantitative representation of the gesture's strength. %For this task, force estimation is not strictly necessary, as the goal is to compare gesture intensities relative to each other rather than to determine an absolute measure of intensity.
\end{itemize}

The primary goal of this work is to demonstrate the feasibility and potential of a vision-based soft-material controller for gesture detection. The use of an elastic, deformable medium introduces unique challenges, including non-linear deformation behaviors, complex optical patterns, and the lack of prior methodologies or datasets. To address these challenges, we combine event-based feature tracking methods and simple rule-based techniques.

\section{Methodology}
%Detailed description of the proposed method, algorithm, or system.
%Include diagrams, flowcharts, or pseudo-code to explain the approach.
%Break down complex methods into subcomponents for clarity.
%Theoretical analysis, proofs, or derivations (if applicable).

\begin{figure*}[t]
  \centering
  \includegraphics[width=\textwidth]{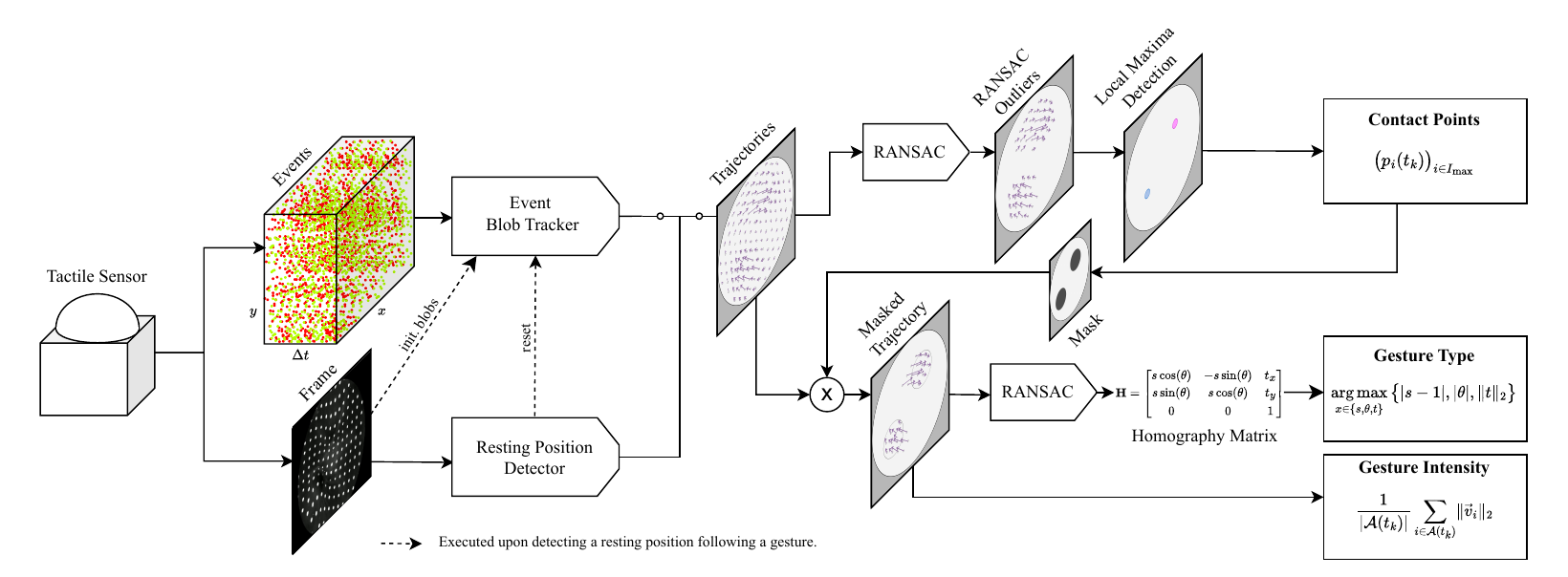}
  \caption{Complete gesture detection framework. Our method leverages marker displacement data to localize contact points, classify gestures and estimate their intensity. Additionally, a resting position detection is performed on the APS frames.}
  \Description{Fully described in caption.}
  \label{fig:gesture_detection_architecture}
\end{figure*}

The complete gesture detection pipeline is illustrated in Figure~\ref{fig:gesture_detection_architecture}. Due to the simplicity of our visual scene (white markers, black background and constant illumination) and to the difficulties in generating large labeled datasets, rule-based methods are particularly emphasized. Our method leverages marker displacement data to infer contact points by identifying local maxima in the displacement fields. The gesture type is estimated by computing a homography matrix used as a classifier and the gesture intensity by averaging the displacement around contact points. Sections~\ref{subsec:marker_tracking}-~\ref{subsec:resting_position} provide a detailed overview of the key algorithmic components.

\footnotetext{The \textit{Twist} gesture is further split into two types: \textit{Clockwise Twist}
(-) and \textit{Counter-Clockwise Twist} (+).}

\subsection{Marker Tracking}\label{subsec:marker_tracking}
The displacements of the markers provide a direct approximation of the gel strain field \cite{of_mlp_tactile_sensor}, offering valuable information for subsequent analysis. We estimate marker displacements by tracking their positions, using the events captured by the EVS and relying on the asynchronous event-blob tracking method introduced by~\cite{AEB_Tracker}, which, by using an Extended Kalman Filter (EKF) to track the state of each blob, surpasses other real-time event-tracking state-of-the-art methods \cite{ACE_event_tracking, HASTE_event_tracking} in both speed and accuracy.

An event-blob refers to a spatio-temporal Gaussian model that describes the likelihood of event occurrences. Let $N$ be the number of markers such that each marker is assigned to an event-blob and $i \in \llbracket 1, N \rrbracket$. The position $p_\varepsilon$ of an event $\varepsilon$ at timestep $t_k$ caused by the $i$-th event-blob of size $\Lambda_i(t_k)\in \mathbb{R}^2_+$ and position $p_i(t_k)\in \mathbb{R}^2$ has the following normal probability distribution:

\begin{equation}
    p_\varepsilon \sim \mathscr{N}(p_i(t_k), \Lambda_i(t_k)^2).
\end{equation}

With this distribution assumption, the state of the blobs (position, velocity and shape) can be estimated with the upcoming events. For each event, the nearest blob's state is updated if its distance falls under a given threshold specified in \cite{AEB_Tracker}. 

Building on \cite{AEB_Tracker}, we initialize blobs positions and sizes on the APS frame recieved at $t_0$, by thresholding the image and extracting white connected regions. Additionally, we introduce the assumptions that all of our markers are circular \big($\left(\Lambda_i(t_k)\right)_x = \left(\Lambda_i(t_k)\right)_y, \; \forall i \in \llbracket 1, N \rrbracket$\big) and have a fixed size ($\Lambda_i(t_k) = \Lambda \in \mathbb{R}^2_+, \; \forall i \in \llbracket 1, N \rrbracket$). By presuming a constant blob size, the EKF from \cite{AEB_Tracker} is linearized, reducing update time. These assumptions are justified by the fact that the markers are dots, and their sizes vary only slightly with respect to the image space. This asynchronous method enables us to track the $N = 177$ markers on the silicone surface in real time. In our method, we represent $\vec{v}_i(t_k)$, the $i$-th marker displacement from $t_0$ to $t_k$, as a linear displacement: $\vec{v}_i(t_k) = p_i(t_k) - p_i(t_0), \; \forall i \in \{1, \dots, N\}$. Here, $t_0$ denotes a timestep where the gel is in resting position, ensuring $\big(\vec{v}_i(t_k)\big)_{i \in \{1, \dots, N\}}$ captures the gel's strain field. We find this linear representation sufficient to detect gesture types, intensities, and contact points effectively. 

\subsection{Contact Point Detection}
\label{contact_point_detection}
 A contact point is the location where the silicone gel experiences the greatest deformation caused by a finger’s interaction. The first step in contact point detection involves identifying markers covered by the user's fingers. When fingers are applied to the gel, they create localized, non-uniform deformations around the contact points (cf. Figure~\ref{fig:gesture_types}), which are reflected in the displacement patterns of the markers. To separate these localized displacements from the overall deformation field, we utilize the RANSAC algorithm \cite{RANSAC} to estimate a homography matrix based on the displacements of the markers, $\big(\vec{v}i(t_k)\big){i \in {1, \dots, N}}$. The finger-induced localized deformations deviate significantly from the global deformation pattern represented by the homography matrix, causing them to be classified as outliers by RANSAC.

RANSAC's outliers are sensitive to the reprojection threshold parameter, which defines the maximum allowable pixel distance between observed and predicted points for a data point to be considered an inlier. With a fixed reprojection threshold, the number of outliers can vary significantly with gesture intensity: smaller displacements yield fewer outliers, while larger displacements produce more. To ensure a consistent subsample size regardless of gesture intensity, we adopt a dynamic reprojection threshold that scales with the average displacement magnitude:

\begin{equation} \texttt{reprojThreshold} = a \cdot \sum_{i=1}^N \frac{\lVert \vec{v}_i(t_k) \rVert_2}{N} \end{equation}
where $a \in \mathbb{R}_+$ is a scalar hyperparameter. After identifying the non-linear displacement subsample $\mathcal{S}(t_k)$, local maxima are detected by identifying markers with the highest displacement relative to their neighbors within a specified radius $r \in \mathbb{R}_+$. Let the neighborhood of the $i$-th marker at timestep $t_k$ be defined by:

\begin{equation}
    \mathcal{N}(i, t_k) = \{ j \mid \| p_i(t_k) - p_j(t_k) \|_2 \leq r, \; \forall j \in \mathcal{S}(t_k)\},
\end{equation}
then, the indices of the markers corresponding to a local peak displacement are defined by:

\begin{equation}
I_{\text{max}}(t_k) = \{ i \mid \lVert \vec{v}_i(t_k) \rVert_2 \geq \lVert \vec{v}_j(t_k)\rVert_2, \; \forall j \in \mathcal{N}(i) \}.
\end{equation}

The contact points at timestep $t_k$ are then defined as the positions of the markers corresponding to these local maximum displacements. 

\begin{equation}
    \text{Contact Points} = \big( p_i(t_k)\big)_{i \in I_{\text{max}}}
\end{equation} 

\subsection{Gesture Classification and Intensity Prediction}
To classify the type and estimate the intensity of each gesture, we analyze the trajectories of markers around the contact points. Let $\mathcal{A}(t_k)$ denote the set of markers whose distance from a contact point is inferior to $r$ at timestep $t_k$. 

To detect the gesture type, we leverage the observation that each gesture of interest resembles a simple transformation when analyzing the displacement of markers of $\mathcal{A}(t_k)$.

\begin{itemize}[label=\textbullet, left=10pt]
    \item A \textit{Twist} gesture (resp. clockwise / counter-clockwise) corresponds to a rotation (resp. negative / positive) $\theta \in \mathbb{R}^\ast$.
    \item A \textit{Pinch} gesture corresponds to a scale decrease $s \in [0,1[$.
    \item A \textit{Zoom} gesture corresponds to a scale increase $s > 1$.
    \item A \textit{Push} gesture corresponds to a translation $t = (t_x, t_y)^\top \in \mathbb{R}^2$.
\end{itemize}

By constraining a transformation to translation, rotation around the center, and scaling from the center, we can represent it using a homography matrix $\mathbf{H}$ of the following form:

\begin{equation}
    \mathbf{H} = \begin{bmatrix}
s \cos(\theta) & -s \sin(\theta) & t_x \\
s \sin(\theta) & s \cos(\theta) & t_y \\
0 & 0 & 1
\end{bmatrix},
\end{equation}

Under these constraints, the mapping $s, \theta, t \to \mathbf{H}(s, \theta, t)$ is injective and the transform parameters can be directly recovered as $s = \sqrt{\mathbf{H}_{11}^2 + \mathbf{H}_{21}^2}, \;
\theta = \text{atan2}\left(\mathbf{H}_{21}, \mathbf{H}_{11}\right)$ and $
t = \left( \mathbf{H}_{13}, \mathbf{H}_{23} \right)^\top$. By approximating the transformation origin as the average position of the contact points, we compute a homography matrix from the displacement vectors of $\mathcal{A}(t_k)$ using RANSAC, constrained to four degrees of freedom. This homography matrix acts as a classifier for gesture types by identifying the dominant simple transformation (translation, scaling, or rotation) within the observed motion at timestep $t_k$:

\begin{equation}
    \text{Gesture Type} \leftarrow \underset{x \in \{s, \theta, t\}}{\arg\max} \, \big\{ |s-1|, |\theta|, \lVert t \rVert_2 \big\}.
\end{equation}

To assess gesture intensity, which quantifies the force applied by the user, the intensity metric should positively correlate with the displacement of the markers in $\mathcal{A}(t_k)$. Therefore, at each timestep $t_k$, we measure the intensity of a gesture as the average displacement of markers within $\mathcal{A}(t_k)$:

\begin{equation} \text{Gesture Intensity} = \frac{1}{|\mathcal{A}(t_k)|} \sum_{i \in \mathcal{A}(t_k)} \lVert \vec{v}_i(t_k) \rVert_2, \end{equation}

This measurement provides a simple linear correlation between displacement and intensity. However, alternative intensity profiles can be employed to better suit specific applications. For instance, a quadratic or more complex non-linear profile could emphasize greater displacements or introduce distinct scaling behaviors.

\subsection{Resting Position Detection}\label{subsec:resting_position}
\label{resting_position}
Over time, the quality of marker tracking can deteriorate during a gesture due to sensor noise artifacts and obstructions (when markers move outside the camera’s field of view or become occluded by the user’s fingers). To maintain tracking quality, it is beneficial to frequently reset the tracker by reinitializing the markers positions. The ideal moment for resetting occurs during a resting position, as there is no marker displacement or gesture activity.

Since the EVS produces only noise-related events when the gel is at rest, resting position detection utilizes APS frames, which are better suited for static scenarios. Operating in parallel to the main pipeline, the resting position detector compares the current APS image to a reference resting position image captured at pipeline launch.

We use the Chamfer distance \cite{ChamferDistance} as the similarity metric, which calculates the distance between two point clouds by summing the distances to the nearest points in each set. Formally, let $\mathcal{P}$ and $\mathcal{Q}$ be two sets of points; the Chamfer distance is defined as:

\begin{equation}
    d_{\text{Chamfer}}(\mathcal{P}, \mathcal{Q}) = \sum_{p \in \mathcal{P}} \min_{q \in \mathcal{Q}} \|p - q\|^2 + \sum_{q \in \mathcal{Q}} \min_{p \in \mathcal{P}} \|p - q\|^2.
\end{equation}

In our case, the point sets $\mathcal{P}$ and $\mathcal{Q}$ represent the pixels corresponding to the markers (i.e. white pixels). The Chamfer distance effectively detects resting positions by measuring the alignment between current and initial marker positions. Practically, a static threshold on the Chamfer distance determines whether the gel is in a resting state.

\section{Experiments and Results}
%Setup: Description of the experimental setup, including hardware, datasets, and metrics.
%Baseline Comparisons: Compare the proposed method with state-of-the-art techniques.
%Results: Quantitative results (e.g., tables, graphs) and qualitative results (e.g., visualizations, robot demonstrations).
%Ablation Studies: Analysis of individual components or parameters to evaluate their contributions.
%Discussion: Interpretation of results, strengths, and any observed limitations.

\subsection{Creation of the Gesture Detection Dataset}
\label{dataset_creation}
\begin{figure*}[htbp]
    \centering
    \begin{minipage}{0.45\textwidth}
        \centering
        \includegraphics[width=\columnwidth]{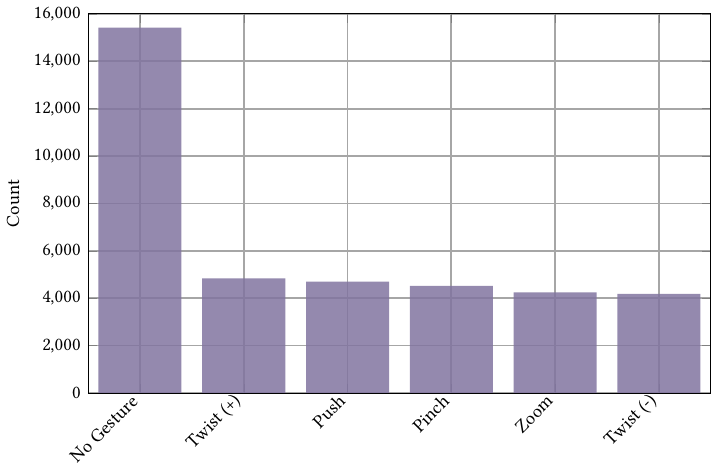}
        \caption{Gesture type labels distribution.}
        \Description{About 15000 frames are labeled as No Gesture. Additionally, there are between 4000 and 5000 frames for each gesture type (Push, Pinch, Clockwise Twist, Counter-Clockwise Twist and Zoom).}
        \label{fig:class_labels_barplot}
    \end{minipage}
    \hfill
    \begin{minipage}{0.45\textwidth}
        \centering
        \includegraphics[width = \columnwidth]{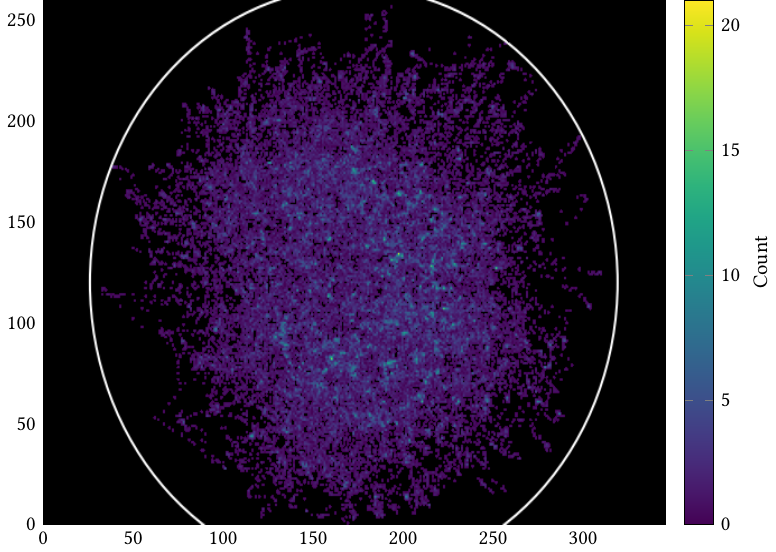}
        \caption{Contact point labels image distribution. White circle marks the gel contour. There is a slight rightward bias, likely due to all users being right-handed.}
        \Description{There are observations approximately everywhere on the except on the edges of the gel. Most contact points are concentrated around the center, with a slight bias toward the right.}
        \label{fig:2d-point-distribution}
    \end{minipage}
\end{figure*}

\begin{figure*}
    \centering
    \includegraphics[width=\textwidth]{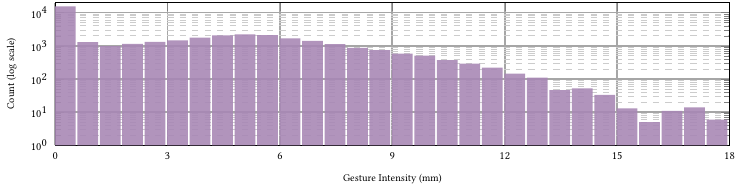}
    \caption{Gesture intensity labels distribution. Each intensity bin has a size of 0.6 mm. \textit{No Gesture} observations are excluded from this bar plot.} 
    \Description{Most of the labeled intensities lie between 0 millimeters and 6 millimeters, then gradually decreases until 18.1 millimeters.}
    \label{fig:intensity_labels_barplot}
\end{figure*}

\begin{figure*}[htbp]
    \centering
    \begin{minipage}{0.6\textwidth}
    \centering
    \includegraphics[width=\textwidth]{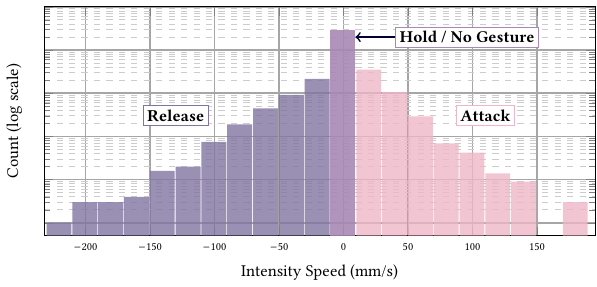}
    \caption{Gesture speed distribution. The gesture speed is calculated as the instantaneous intensity change between consecutive frames.}
    \Description{Negative speeds indicate a release phase and positive speeds indicate an attack phase. A null speed indicates either a holding phase or a resting position. The gesture speed distribution is clearly Gaussian, centered on 0, spanning from -220 millimeters per second to 190 millimeters per second.}
\label{fig:speed_distribution}
    \end{minipage}
    \hfill
    \begin{minipage}{0.38\textwidth}
        \centering
        \includegraphics[width = \textwidth]{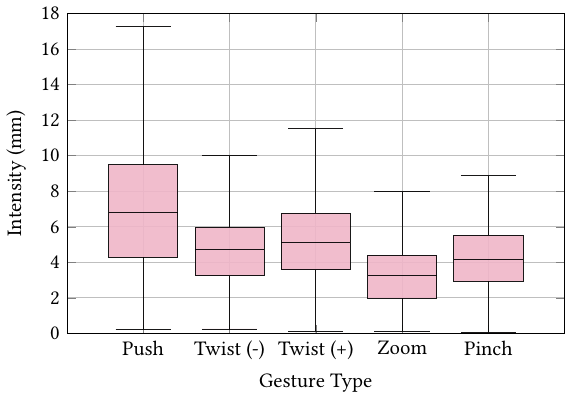}
        \caption{Gesture intensity labels per gesture type box plot.}
        \Description{The average intensities for all gesture types fall within the range of 3 to 7 millimeters. Among the gestures, the Zoom gesture exhibits the lowest average intensity, spanning from 0 to 8 millimeters. In contrast, the Push gesture shows the highest average intensity and ranges from 0.1 to 17.5 millimeters.}
        \label{fig:intensity_boxplot}
    \end{minipage}
\end{figure*}

To the best of our knowledge, no established benchmarks currently exist for gesture detection on vision-based soft-material controllers, whether using frame-based or event-based vision. To evaluate the performance of our gesture detection pipeline, we have recorded and labeled a gesture detection dataset. This dataset contains \SI{25}{\minute} of gesture recordings from 5 different users interacting with NeuroTouch. Each user has performed around \SI{1}{\minute} of each basic gesture type (\textit{Push}, \textit{Pinch}, \textit{Zoom}, \textit{Clockwise Twist}, and \textit{Counter-Clockwise Twist}) at various locations on the silicone gel (cf. Figure~\ref{fig:2d-point-distribution}), with varying intensities (local deformations up to \SI{1.8}{\centi\meter} displacement, cf. Figure~\ref{fig:intensity_labels_barplot}), speeds (as high as \SI{210}{\milli\meter\per\second}, cf. Figure~\ref{fig:speed_distribution}) and number of fingers (up to 3). Each gesture comprises three distinct phases:
\begin{itemize}[label=\textbullet, left=10pt]
\item \textbf{Attack}: The gel begins to deform as external force is applied by the user, with the deformation increasing progressively in intensity.
\item \textbf{Hold}: The gel maintains its maximum deformation as the applied force is sustained.
\item \textbf{Release}: The gel gradually returns to its original shape as the applied force decreases and is fully removed by the user.
\end{itemize}

In this dataset, an observation is defined as the gesture type, gesture intensity, and contact points recorded at a specific timestamp. Importantly, an observation represents a single moment within a gesture rather than an entire gesture itself. Labels are manually annotated on the $\num{37912}$ APS frames of the dataset. The contact points are visually estimated by selecting the marker closest to the deformation peak caused by the finger. Given that a finger can cover several markers on the gel, visually pinpointing the exact contact point can be challenging. As a result, the contact point measurement has a minimum uncertainty of \SI{4}{\milli\meter}, corresponding to the spacing between markers. Gesture intensity is estimated by calculating the average displacement of the labeled contact points from the start of the gesture. To compare predictions and labels at specific timestamps, we use the most recent labeled gesture type and linearly interpolate the contact points localizations and intensities based on the preceding and subsequent labeled data. Statistics on the dataset can be found in Figures~\ref{fig:class_labels_barplot}-~\ref{fig:intensity_boxplot}.

\subsection{Metrics}
For gesture type classification, standard metrics such as precision, recall, and F1-score are used for each class and global accuracy is measured. Gesture intensity is assessed using the mean absolute error (MAE) between predictions and ground truth on labels where a gesture is performed. For contact point detection, two metrics are used:
\begin{itemize}[label=\textbullet, left=10pt]
\item \textbf{Average Euclidean Distance Error}: Measures the average distance between each predicted contact point and its nearest ground truth contact point. In cases where the number of predicted contact points exceeds the number of ground truth contact points, the furthest predicted points from any ground truth point are discarded.
\item \textbf{Contact Point Count Accuracy}: Measures the accuracy of the predicted number of contact points.
\end{itemize}

In addition to these metrics, we conduct an analysis of our pipeline's accuracy with respect to gesture intensity, along with a runtime performance study that evaluates the execution time of each component.

\subsection{Results}
\label{subsec:results}

\begin{figure}
    \centering
    % First subfigure
    \begin{subfigure}{0.23\textwidth}
        \includegraphics[trim={25pt 25pt 25pt 25pt}, clip, width=\textwidth]{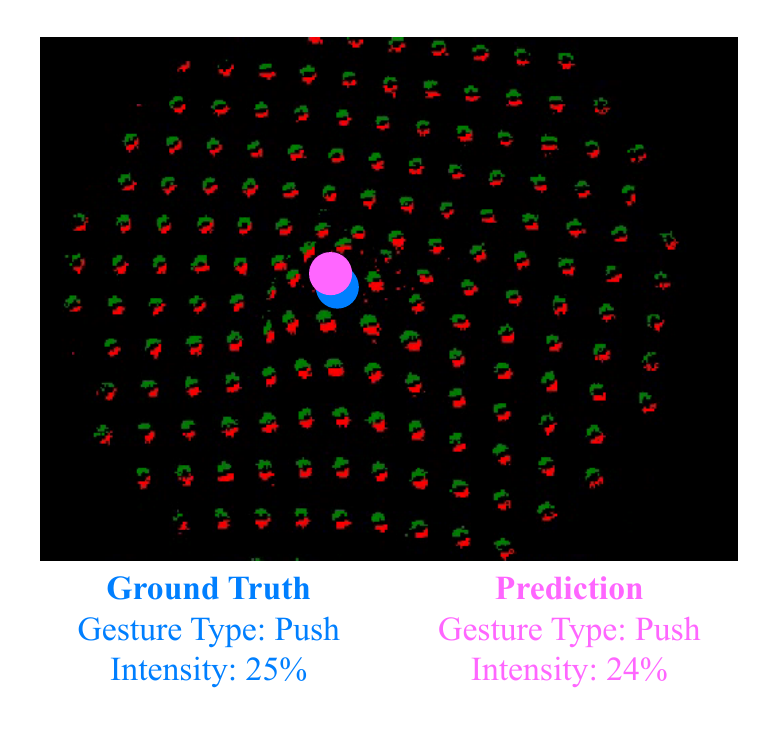}
    \end{subfigure}
    % Second subfigure
    \begin{subfigure}{0.23\textwidth}
        \includegraphics[trim={25pt 25pt 25pt 25pt}, clip, width=\textwidth]{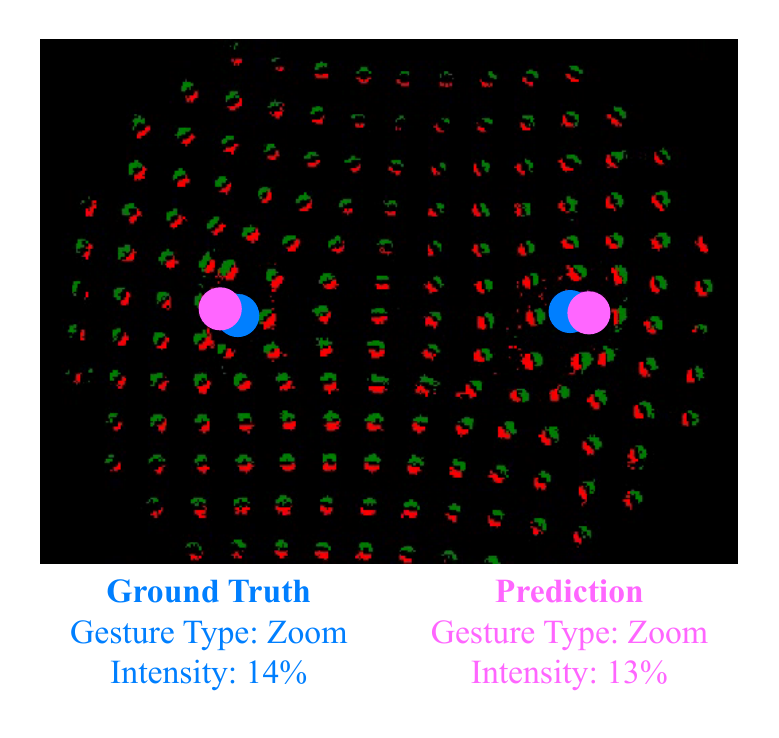}
    \end{subfigure}
    % Third subfigure
    \begin{subfigure}{0.23\textwidth}
        \includegraphics[trim={25pt 25pt 25pt 25pt}, clip, width=\textwidth]{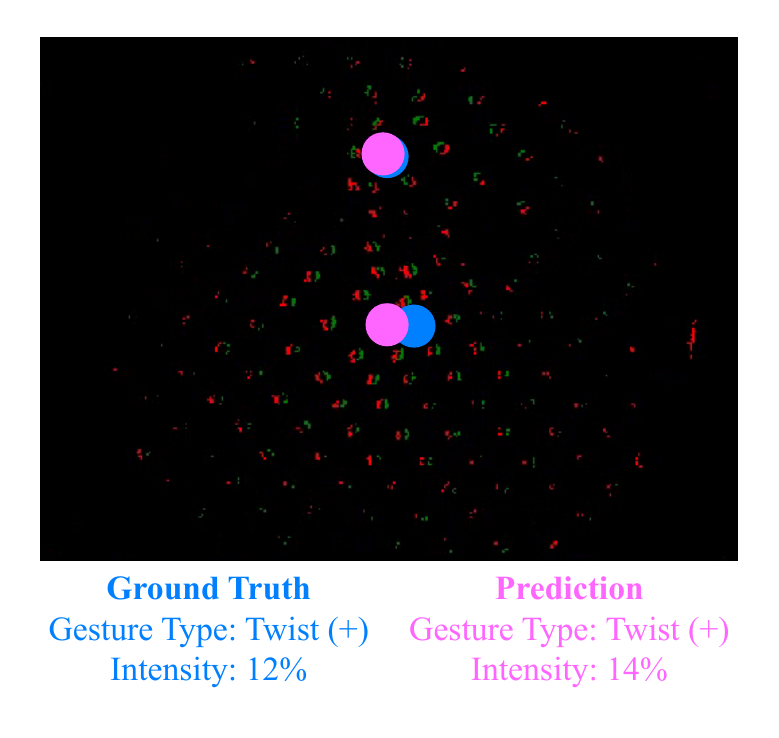}
    \end{subfigure}
    % Fourth subfigure
    \begin{subfigure}{0.23\textwidth}
        \includegraphics[trim={25pt 25pt 25pt 25pt}, clip, width=\textwidth]{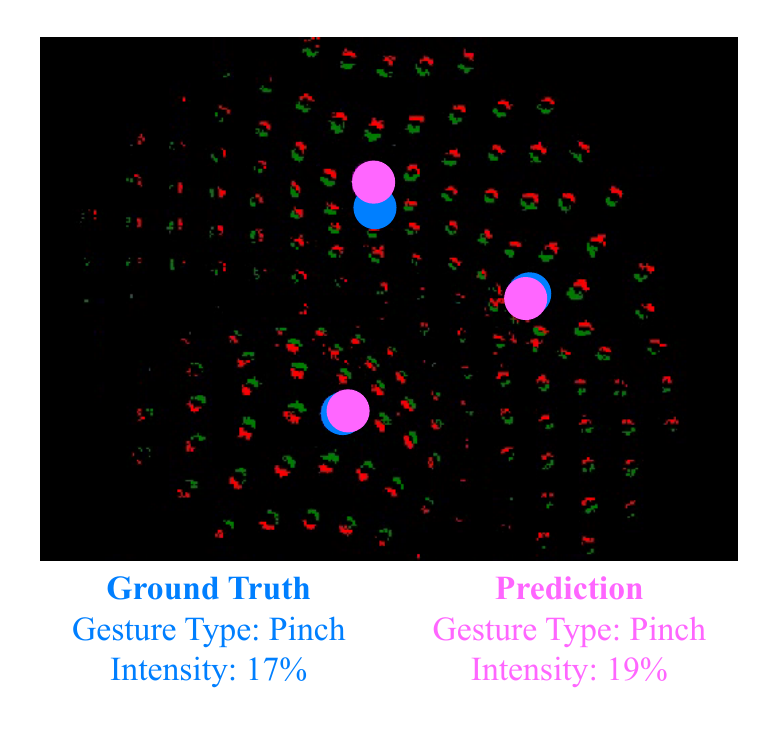}
    \end{subfigure}
    \caption{Prediction examples from the dataset. Pink (resp. blue) dots represent the predicted (resp. labeled) contact points on the events image space. Intensity is scaled by the radius of the gel (30 mm).}
    \Description{The images compare the predictions from the ground truth on several examples. Ground truth and predicted contact points and intensities are close to each other and gesture type is the same.}
    \label{fig:labels_vs_preds}
\end{figure}

The pipeline is executed on the complete dataset and the results are evaluated against the labeled ground truth. For the hyperparameters, we select a radius $r$ of $ 30 \; \text{pixels} \approx$ \SI{12}{\milli\meter}, which is the typical radius of a fingertip. The dynamic reprojection threshold coefficient $a$ is set to $0.6$. To ensure the reliability of local maxima detection and mitigate the impact of outliers, we require a minimum number of neighbors, $N_{min} = 4$, such that $|\mathcal{N}(i,t_k)| \geq N_{min}, \; \forall i,t_k$. Additionally, the Chamfer distance static threshold used to detect a resting position is set to $2.5$. Prediction examples are displayed in Figure~\ref{fig:labels_vs_preds}.

\subsubsection{Gesture Detection Metrics}

\begin{figure}[htbp!]
\centering
\includegraphics[width=0.9\columnwidth]{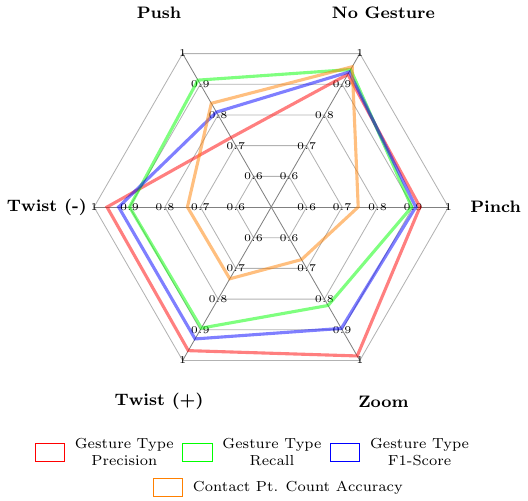}
\caption{Gesture type and contact point count classification metrics}
\Description{Fully described in text.}
\label{fig:gesture_type_metrics}
\end{figure}

\begin{figure*}
    \centering
    \includegraphics[width = 0.6\textwidth]{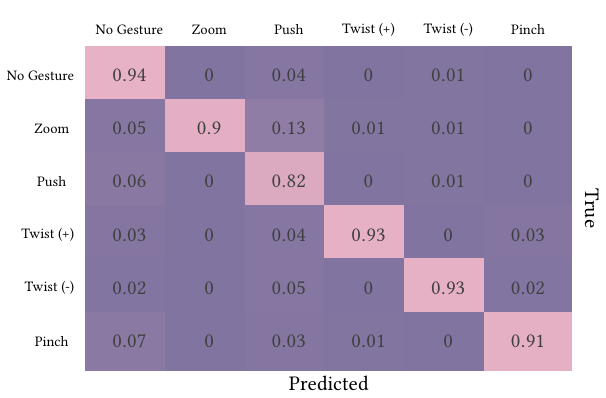}
    \caption{Balanced gesture type confusion matrix, averaged between raw-normalized and column-normalized values.}
    \Description{The confusion matrix shows the classification performance for No Gesture, Zoom, Push, Counter-Clockwise Twist, Clockwise Twist, and Pinch. Most predictions align with the true classes and confusions are minor. The biggest confusion is the Zoom gesture, which is sometimes misclassified as a Push with a value of 0.13.}
    \label{fig:classification_matrix}
\end{figure*}

Figure~\ref{fig:gesture_type_metrics} summarizes the precision, recall, and F1-score for each gesture type, complemented by the contact point count accuracy. A gesture type confusion matrix is also provided in Figure~\ref{fig:classification_matrix}. Our pipeline achieves an average gesture type accuracy of \SI{91.12}{\percent}, with excellent precision for multi-finger gestures (\textit{Twist}, \textit{Pinch}, \textit{Zoom}) but lower precision and higher recall for one-finger gestures (\textit{Push}). Additionally, our pipeline achieves an average contact point count accuracy of \SI{83.22}{\percent}, showing better performance for resting positions or one-finger gestures compared to the ones involving multiple fingers. This disparity in gesture type and contact point count accuracy arises because multi-finger gestures often exhibit varying intensities across individual fingers. When one finger dominates in displacement, a \textit{Push} is more likely to be predicted at the start of the gesture, before the true multi-finger gesture is fully formed. Furthermore, resting position detection using Chamfer distance achieves a balanced F1-score of \SI{94.03}{\percent}, with well-aligned precision and recall.

\begin{figure}[htbp!]
    \centering
    \includegraphics[width = \columnwidth]{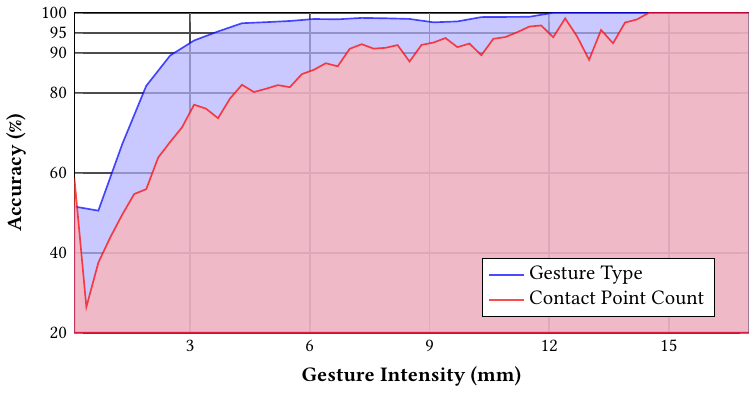}
    \caption{Gesture type and contact point count accuracy as a function of ground truth intensity.}
    \Description{The intensity is provided in millimeters and can be interpreted as the average gel surface displacement around the user's fingers. Observations labeled as No Gesture have a null intensity and are excluded the results used to draw this bar plot. The gesture type accuracy gradually increases over the intensity, reaching 90 percent at 3 millimeters, 95 percent at 4.2 millimeters and 100 percent at 12 millimeters. The curve shows that contact point count accuracy reaches 80 percent at approximately 6.1 millimeters and 90 percent at around 7.0 millimeters of gesture intensity, then continues to increase rapidly, reaching near-perfect accuracy at approximately 14.5 millimeters, where it plateaus.}
    \label{fig:accuracy_vs_intensity}
    \end{figure}

The pipeline gesture recognition performance is also heavily influenced by the gesture intensity. Figure~\ref{fig:accuracy_vs_intensity} shows a positive correlation between the gesture type and contact point count accuracy relative to ground truth intensity. Our method achieves a gesture type accuracy greater than \SI{90}{\percent} when the average marker displacement exceeds only \SI{2.5}{\milli\meter}. Notably, the pipeline reaches a \SI{100}{\percent} accuracy for intensities larger than \SI{12.1}{\milli\meter}. These results emphasize that during a gesture, the detection likelihood is lower at the start, increases during the attack phase, remains high throughout the holding phase, and decreases during the release phase.

\begin{figure}[htbp]
\centering
\includegraphics[width=\columnwidth]{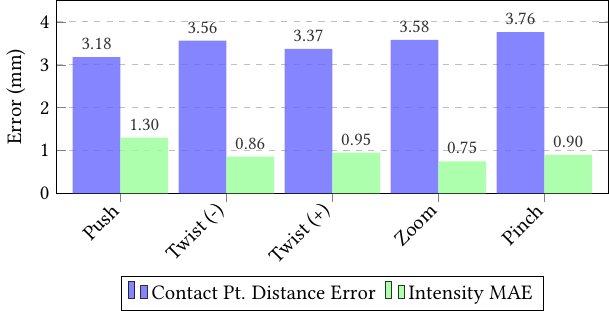}
\caption{Contact point localization and intensity error across gesture types.}
\Description{Fully described in text.}
\label{fig:dist_vs_gesture}
\end{figure}

Figure~\ref{fig:dist_vs_gesture} shows the contact point Euclidean distance error and gesture intensity MAE across gesture types. Our pipeline achieves an average contact point distance error of \SI{3.41}{\milli\meter}, with consistent results across all gesture types, all of which fall below the \SI{4}{\milli\meter} contact point localization labeling uncertainty. This indicates that the pipeline accurately estimates the finger positions on the gel. Reducing this uncertainty, through higher marker resolution or more precise labeling methods, could further validate the model's precision. For gesture intensity prediction, the pipeline achieves a sub-millimeter MAE of \SI{0.96}{\milli\meter}, with slightly higher error for \textit{Push} gestures and lower error for \textit{Zoom} gestures. This arises from the intensity error scaling with gesture displacement: \textit{Push} involves slightly larger displacements, while \textit{Zoom} usually involves smaller ones (cf. Figure~\ref{fig:intensity_boxplot}). This is primarily because the extensor muscles (used for zooming) are weaker than the other muscles in the hand and forearm \cite{finger_flexion_vs_extension}, making the gesture physically harder to perform.

\subsubsection{Runtime Performance}
\label{runtime_performance}
For our runtime study, we executed our pipeline on a CPU-only setup utilizing two cores of an Intel(R) Xeon(R) W-2235 processor with a clock speed of \SI{3.80}{\giga\hertz}. One core is dedicated to resting position detection using frames, with processing frequency constrained by the camera APS frame rate (\SI{25}{\hertz}). The second core handles the gesture detection. Events are currently processed in \SI{10}{\milli\second} non-overlapping windows, resulting in a gesture detection frequency of \SI{100}{\hertz}.

Table~\ref{tab:runtime_performance} summarizes the runtime performance of the pipeline components.  
Our asynchronous blob tracker has an average throughput of \num{4.485} MegaEvents per second, and processes a \SI{10}{\milli\second} event batch on average in \SI{2}{\milli\second}, demonstrating the tracker's ability to operate in real time. Contact point detection, gesture classification, and intensity estimation take an additional \SI{2.28}{\milli\second} to process the marker trajectories. Thus, our gesture detection pipeline could theoretically operate with \SI{3}{\milli\second} event batches, achieving a frequency of \SI{333}{\hertz}. Resting position detection has an average processing time of \SI{6.54}{\milli\second} per frame, indicating that it could support real-time operation at frame rates of up to \SI{152}{\hertz}.

% Mention the GFLOPS estimation? Is our model acutally low-power for embedded systems that use microcontrollers?

\section{Discussion}
%Broader implications of the results.
%Challenges and limitations of the approach.
%Potential applications or extensions.

This work highlights the potential of optical-based tactile sensing to transform human-computer interaction. By enabling precise gesture classification and contact point detection on a deformable silicone surface, our approach introduces a novel interface paradigm that integrates soft tactile input with digital responsiveness. The high accuracy and low latency demonstrated in our experiments suggest its applicability in video games, AR/VR, accessible devices, and more.

\begin{table}[t!]
\centering

\caption{Runtime performance}
\label{tab:runtime_performance}
\begin{tabular}{cccc}
\toprule
\textbf{Core} & \textbf{Component}                           & \textbf{Avg (\small{\si{\milli\second}})} & \textbf{Std (\small{\si{\milli\second}})} \\ 
\midrule
\multirow{4}{*}{\textbf{1st}} 
              & Marker Tracking \footnotesize{(\SI{10}{\milli\second} event batch)}          & 1.97 & \textbf{2.22}     \\ 
              & Contact Point Detection             & \textbf{1.98}     & 0.74              \\ 
              & Gesture Type / Intensity              & 0.30              & 0.10              \\ 
              & \textbf{Total}                               & \textbf{3.63}     & \textbf{2.66}     \\ 
\midrule
\textbf{2nd} 
              & Resting Position Detection                          & 6.54              & 1.67              \\
\bottomrule
\end{tabular}
\end{table}

Future work could focus on improving the device's sensitivity to normal forces. One potential approach is to estimate the depth of the markers to provide 3D trajectories, a feature that our pipeline can readily accommodate with minor modifications. Additionally, increasing the marker resolution might also improve normal forces detection, though this could require optimizing runtime efficiency and adjusting the pipeline accordingly. Another promising direction involves addressing power consumption by relying solely on the event-based sensor. This could be achieved by using flickering LEDs to generate event frames, effectively replacing the APS frame without compromising the tracker performance \cite{AEB_Tracker}.

\begin{figure*}
    \centering
    \includegraphics[width=0.7\textwidth]{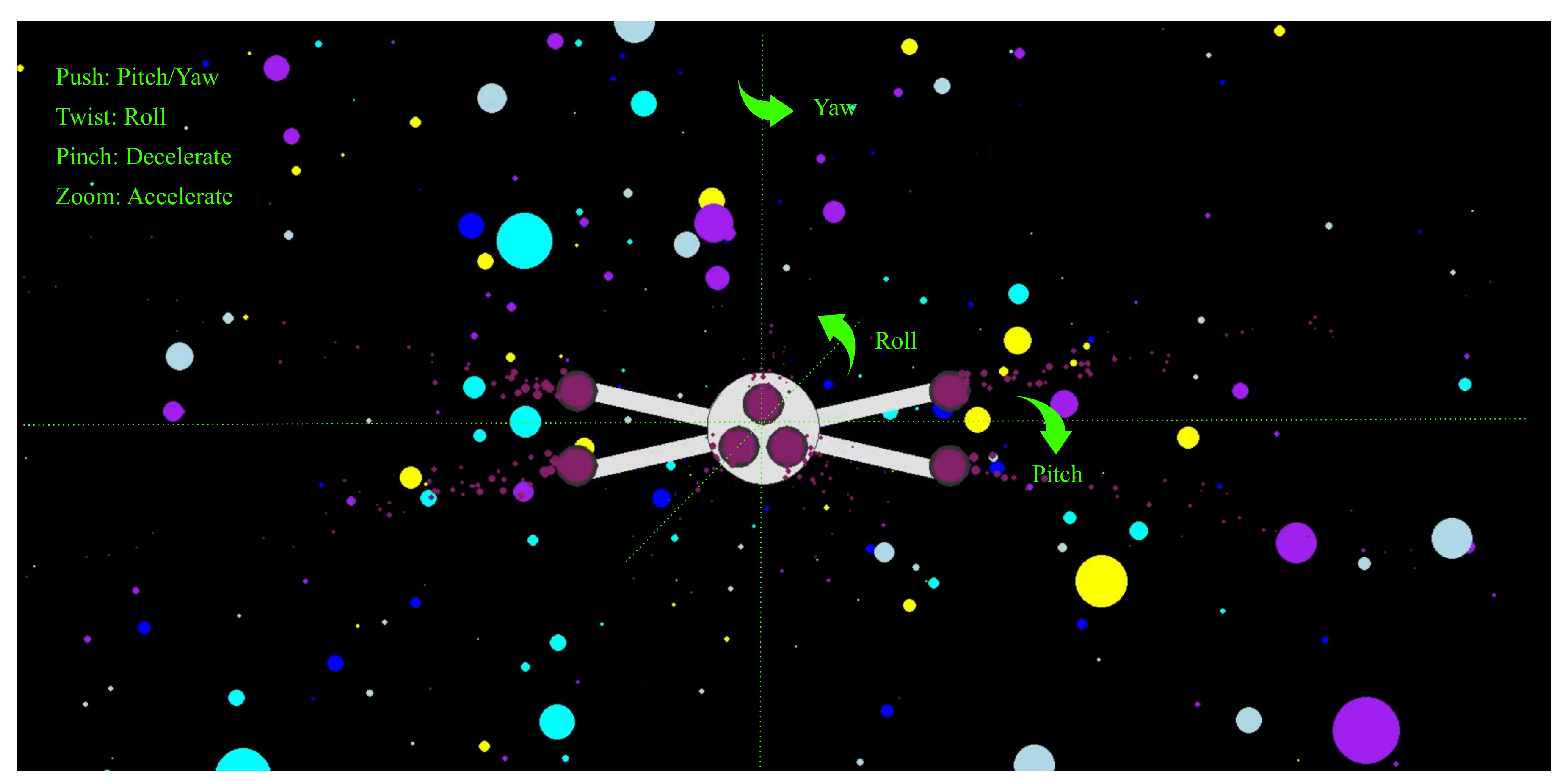}
    \caption{Snapshot of a simple spacecraft video game using NeuroTouch. Gestures on the soft material are used to control the spaceship with adaptive intensity. Complete video: Coming Soon}
    \Description{A Push gesture in a given direction allows the spacecraft to pitch or yaw, while twisting the gel enables it to roll. Additionally, zooming and pinching gestures allow the spacecraft to accelerate and decelerate, respectively.}
    \label{fig:video_game_snapshot}
\end{figure*}

Our framework is highly scalable, adaptable to various silicone shapes, and compatible with existing input devices such as gamepads, either to enhance functionality or to serve as a standalone tactile controller. A video game application example of NeuroTouch is illustrated in Figure~\ref{fig:video_game_snapshot}. The diversity of controls achievable with just two-finger gestures also makes it particularly valuable for accessible devices for individuals with impaired hand function. We can also expand the range of gestures for more tailored interactions based on application. Additionally it is possible to develop smaller NeuroTouch prototypes, as our method is equally effective with compact neuromorphic cameras and lenses. As tactile interfaces evolve, the principles in this study could drive significant innovations in soft robotics and interactive technologies.

\section{Conclusion}
%Summary of the main findings and contributions.
%Brief mention of future research directions.
This work presents a new approach to gesture detection using NeuroTouch, a vision-based soft material controller. Our system leverages a curved silicone gel embedded with markers and a neuromorphic camera to accurately track multi-finger gestures in real time. With a \SI{3.41}{\milli\meter} contact point localization error, \SI{91}{\percent} gesture classification accuracy, and \SI{0.96}{\milli\meter} intensity estimation error on our publicly available dataset, NeuroTouch demonstrates its feasibility for intuitive and expressive interaction paradigms. Future research will explore extending the sensor design, refining the detection pipeline, and expanding the system's utility in diverse environments.

\bibliographystyle{ACM-Reference-Format}
\bibliography{bibliography}

\clearpage

\begin{comment}

\begin{figure*}
    \centering
    % First subfigure
    \begin{subfigure}{0.24\textwidth}
        \includegraphics[trim={25pt 25pt 25pt 25pt}, clip, width=\textwidth]{figures/push_label_vs_pred.pdf}
        \caption{1-finger Push}
    \end{subfigure}
    % Second subfigure
    \begin{subfigure}{0.24\textwidth}
        \includegraphics[trim={25pt 25pt 25pt 25pt}, clip, width=\textwidth]{figures/zoom_label_vs_pred.pdf}
        \caption{2-Finger Zoom}
    \end{subfigure}
    % Third subfigure
    \begin{subfigure}{0.24\textwidth}
        \includegraphics[trim={25pt 25pt 25pt 25pt}, clip, width=\textwidth]{figures/twist_label_vs_pred.pdf}
        \caption{2-Finger Twist (+)}
    \end{subfigure}
    % Fourth subfigure
    \begin{subfigure}{0.24\textwidth}
        \includegraphics[trim={25pt 25pt 25pt 25pt}, clip, width=\textwidth]{figures/pinch_label_vs_pred.pdf}
        \caption{3-Finger Pinch}
    \end{subfigure}
    \caption{Prediction examples from the dataset. Pink (resp. blue) dots represent the predicted (resp. labeled) contact points. Intensity is scaled by the radius of the gel (30 mm).}
    \Description{The images compare the predictions from the ground truth on several examples. Ground truth and predicted contact points and intensities are close to each other and gesture type is the same.}
    \label{fig:labels_vs_preds}
\end{figure*}
    
\end{comment}

\end{document}